\documentclass[12pt, aps,prd,tightenlines,notitlepage,nofootinbib,floatfix]{revtex4-1}
\usepackage[svgnames]{xcolor}
\usepackage[colorlinks=true,urlcolor = violet,linkcolor=blue,citecolor=teal,bookmarks=false]{hyperref}
\usepackage{amsfonts,amsmath,amssymb, bm}
\usepackage[T1]{fontenc} 

\usepackage{dsfont}
\usepackage{lineno}
\usepackage{graphicx} 

\def\Xint#1{\mathchoice
   {\XXint\displaystyle\textstyle{#1}}%
   {\XXint\textstyle\scriptstyle{#1}}%
   {\XXint\scriptstyle\scriptscriptstyle{#1}}%
   {\XXint\scriptscriptstyle\scriptscriptstyle{#1}}%
   \!\int}
\def\XXint#1#2#3{{\setbox0=\hbox{$#1{#2#3}{\int}$}
     \vcenter{\hbox{$#2#3$}}\kern-.5\wd0}}
\def\dashint{\Xint-}

\def\be{\begin{equation}}
\def\ee{\end{equation}}

\def\bp{\begin{pmatrix}}
\def\ep{\end{pmatrix}}

\def\dd{\mathrm{d}}
\def\ii{\mathrm{i}}

\begin{document}
\title{An exact solution to asymptotic Bethe equation}%
\author{Yuan Miao}
\affiliation{Institute for Theoretical Physics, University of Amsterdam, Postbus 94485, 1090 GL Amsterdam, The Netherlands}
\date{\today}

\begin{abstract}
We present an exact solution to the asymptotic Bethe equation of weakly anisotropic Heisenberg spin chain, which is a set of non-linear algebraic equations. The solution describes the low-energy excitations above ferromagnetic ground state with fixed magnetisation, and it has a close relation to generalised Jacobi polynomial. It is equivalent to a generalised Stieltjes problem and in the continuous limit, it becomes a Riemann-Hilbert problem closely related to the finite-gap solutions of classical Landau--Lifshitz field theory.
\end{abstract}

\maketitle

\section{Introduction}
\label{sec:intro}

The study of the emergence of classical dynamics of quantum interacting systems has led to the discoveries of new phenomena recently, especially in the regime of quantum integrable systems. One of the most notable examples is the classical-quantum correspondence of spin transport in the domain-wall quench between quantum anisotropic Heisenberg model and the corresponding classical Landau-Lifshitz field theory~\cite{Gamayun_2019, Misguich_2019}, closely related to semi-classical eigenstates of the spectrum~\cite{YM_classical_quantum}. Moreover, the superdiffusive behaviour of spin transport in quantum isotropic Heisenberg model~\cite{Znidaric_2011, Ljubotina_2017} is attributed to the existence of thermally dressed interacting `giant magnons'~\cite{Ilievski_2018, De_Nardis_2019, Vasseur_2019, De_Nardis_2020, De_Nardis_2020_2}, a manifestation of semi-classical eigenstates as well.

One way to understand the problems above is to use asymptotic Bethe ansatz\cite{Sutherland_1978, Sutherland_1995, Dhar_2000}, i.e. a semi-classical limit of the Bethe ansatz~\footnote{Asymptotic Bethe ansatz here refers to the leading order of Bethe ansatz equations with respect to the effective Planck constant. It is not related to the Bethe--Yang equation in integrable field theories which is usually denoted as thermodynamic Bethe ansatz.}. There are systematic explorations for the solutions to asymptotic Bethe equation of quantum isotropic Heisenberg model~\cite{Kazakov_2004, Bargheer_2008}, motivated as a limit of $\mathcal{N} = 4$ supersymmetric Yang-Mills theory~\cite{Minahan_2003, Kruczenski_2004}. In addition, the results are helpful to obtain the three-point correlation functions in the supersymmetric gauge theories~\cite{Escobedo_2011, Gromov_2012, Kostov_2012_PRL, Kostov_2012}, using a semi-classical limit of the overlaps between the Bethe states.

In this article we derive an exact solution to the asymptotic Bethe equation of quantum weakly anisotropic XXZ model, following the example of Ref.~\cite{Shastry_2001} of the isotropic case.

\section{Asymptotic Bethe ansatz}
\label{sec:ABE}

We focus on the ferromagnetic anisotropic Heisenberg model, also known as XXZ model, and its Hamiltonian can be written in terms of Pauli matrices,
\be 
	H = - \frac{1}{4} \sum_{j=1}^L \left(  \sigma^{\rm x}_j \sigma^{\rm x}_{j+1}  + \sigma^{\rm y}_j \sigma^{\rm y}_{j+1} + \Delta \sigma^{\rm z}_j \sigma^{\rm z}_{j+1} \right) ,
\ee
where we use periodic boundary condition ($\sigma^\alpha_{L+1} = \sigma^\alpha_{1}$), and $\sigma^\alpha_j = \mathbb{I}^{\otimes (j-1)} \otimes \sigma^\alpha \mathbb{I}^{\otimes (L -j)}$. In this article, we concentrate on $\Delta = \cosh \eta > 1$ regime, also known as {\it gapped regime}, due to the property of its antiferromagnetic ground state.

The eigenstates of the Hamiltonian can be characterised by a set of rapidities, i.e. Bethe roots. For each eigenstates with $M$ down spins ($M$ magnons), it can be denoted by $| \{ \vartheta_j \}_{j=1}^M \rangle$, and the Bethe roots $\vartheta_j$ satisfy
\be  
	\left[ \frac{\sin (\vartheta_j + i \frac{\eta}{2})}{\sin (\vartheta_j - i \frac{\eta}{2})} \right]^L
	\prod_{k\neq j}^M \frac{\sin (\vartheta_j - \vartheta_k - i \eta)}{\sin (\vartheta_j - \vartheta_k + i \eta)} = 1 , 
	\label{eq:betheeq}
\ee
i.e. {\it Bethe equations}. They are closely related to the conserved charges associated to the eigenstates. For each magnons, the pseudo-momentum and pseudo-energy in terms of its Bethe root are
\be 
	p (\vartheta) = - \ii \log \frac{\sin ( \vartheta + i \frac{\eta}{2})}{\sin (\vartheta - i \frac{\eta}{2})} , \quad \varepsilon (\vartheta) = \frac{\sin^2 \ii \eta }{\cos 2 \vartheta - \cos \ii \eta } ,
\ee
and the physical momentum and energy of eigenstate $| \{ \vartheta_j \}_{j=1}^M \rangle$ become
\be 
	P = \sum_{j=1}^M p (\vartheta_j ) = - \ii \sum_{j=1}^M \log \frac{\sin ( \vartheta_j + i \frac{\eta}{2})}{\sin (\vartheta_j - i \frac{\eta}{2})}, \quad E = \sum_{j=1}^M \varepsilon (\vartheta_j ) = \sum_{j=1}^M \frac{\sin^2 \ii \eta }{\cos 2 \vartheta_j - \cos \ii \eta } .
\ee

We define the Baxter Q-functions for $| \{ \vartheta_j \}_{j=1}^M \rangle$ as
\be 
	Q (\vartheta) = \prod_{j=1}^M \sin (\vartheta - \vartheta_j ) ,
	\label{eq:Qfunc}
\ee
which is a Laurent polynomial of degree $M$ with variable $t = e^{\ii \vartheta}$. And we could rewrite the Bethe equations~\eqref{eq:betheeq} and take the logarithmic form using shorthand notation $f^{[n]}(\vartheta) = f (\vartheta + \ii n \eta /2)$,
\be 
	\log Q_0^{[+1]} (\vartheta_j) - \log Q_0^{[-1]} (\vartheta_j) = 2 \pi \ii n_j  + \log Q_{j}^{[+2]} (\vartheta_j ) -\log Q_{j}^{[-2]} (\vartheta_j ), \, n_j \in \mathbb{Z} ,
	\label{eq:betheeqlog}
\ee
where we use the following notations
\be 
	Q_0 (\vartheta) = \sin^L \vartheta , \quad Q_j (\vartheta) = \prod_{k\neq j}^M \sin (\vartheta - \vartheta_k ) .
\ee

We use the formalism of asymptotic Bethe ansatz~\cite{Sutherland_1978, Sutherland_1995}, i.e. taking the ``semi-classical limit'' of Bethe equations~\eqref{eq:betheeqlog}~\cite{YM_classical_quantum}, which is equivalent to the weakly anisotropic limit,
\be 
	\eta = \frac{\epsilon \ell}{L} + \mathcal{O} \left( \frac{1}{L^2} \right) , \quad \epsilon = \sqrt{\delta} , \quad \delta = \lim_{a\to 0} \frac{2(\Delta-1)}{a^2} \in \mathcal{O} (1) ,
\ee
where classical period $\ell = L a \in \mathcal{O} (1)$ as system size $L\to \infty$ and lattice spacing $a = \frac{\ell}{L} \to 0$. 

In this limit, we could easily obtain that
\be 
	\log Q_0^{[\pm 1]} (\vartheta_j) = \log Q_0 (\vartheta_j) \pm \frac{\ii \eta}{2} \left. \frac{\dd }{\dd \vartheta}  \log Q_0 (\vartheta) \right|_{\vartheta = \vartheta_j} + \mathcal{O} \left( \frac{1}{L^2} \right) ,
\ee
and 
\be 
	\log Q_j^{[\pm 2]} (\vartheta_j) = \log Q_j (\vartheta_j) \pm \ii \eta \left. \frac{\dd }{\dd \vartheta}  \log Q_j (\vartheta) \right|_{\vartheta = \vartheta_j} + \mathcal{O} \left( \frac{1}{L^2} \right) .
\ee

We define variable $\mu$ as
\be 
	\mu = \frac{\eta}{\ell } \frac{\dd }{\dd \vartheta}  \log Q_0 (\vartheta) = \frac{\epsilon}{\tan \vartheta} ,
\ee
and the Bethe equation~\eqref{eq:betheeqlog} can be recast as
\be 
	\mu_j = \frac{2 \pi}{\ell} n_j - \frac{2}{L} \sum_{k\neq j}^M \frac{\mu_j \mu_k +\delta}{\mu_j - \mu_k} + \mathcal{O} \left( \frac{1}{L} \right) .
	\label{eq:ABE}
\ee
Neglecting the finite-size correction, we conveniently denote Eq.~\eqref{eq:ABE} as {\it asymptotic Bethe equation}. Eq.~\eqref{eq:ABE} can be considered as a generalised version of equations of Stieltjes in random matrix theory~\cite{Shastry_2001}. The remaining part of the article aims at giving an exact solution to the asymptotic Bethe equation~\eqref{eq:ABE} in the simplest case, when all the Bethe roots form a ``huge bound state'' sharing the same mode number $n_1$ and filling fraction $\nu = \frac{M}{L} \in \mathcal{O} (1)$.

We assume that the Bethe roots $\mu_j$ condense around certain finite disjoint segments $\mathcal{C}_j$, and take the continuous limit of Eq.~\eqref{eq:ABE},
\be 
	\frac{\ell \mu}{2} = \pi n_{j} - \ell \dashint_{\mathcal{C}} \dd \lambda\,\frac{\mu \lambda +\delta}{\mu - \lambda}\rho (\lambda) + \mathcal{O} \left( \frac{1}{L} \right)  , \quad \mu \in \mathcal{C}_{j},
\label{eq:contiABE}
\ee
where $\mathcal{C} = \cup_j \mathcal{C}_j$, and density
\be 
	\rho (\mu ) = \lim_{L \to \infty} \frac{1}{L} \sum_{j=1}^M \bm{\delta} (\mu - \mu_j ) .
\ee
Dirac delta function is denoted as $\bm{\delta} (\mu)$. The finite-size (quantum) correction (the $\mathcal{O} (1/L) $ correction in Eq.~\eqref{eq:contiABE}) has been derived in Ref.~\cite{YM_classical_quantum, Beisert_2005}, namely,
\be 
	\frac{1}{L}\pi \rho^\prime (\mu) \ell^{2} (\mu^{2} + \delta)^{2} \coth \left[ \pi \ell (\mu^2 + \delta) \rho (\mu) \right] ,
\label{eq:finite_size_mu}
\ee
which we neglect in this article. As it turns out, when the finite-size correction diverges due to the property of $\coth$ function, the result of this article would require modifications, which is explained in Ref.~\cite{YM_classical_quantum}.

Eq.~\eqref{eq:contiABE} is a Riemann-Hilbert problem, and it coincides with the finite-gap solutions of classical Landau--Lifshitz field theory, which is precisely the semi-classical limit of quantum XXZ model with weak anisotropy. The extensive discussions of the semi-classical quantisation of quantum anisotropic Heisenberg model have been given in Ref.~\cite{YM_classical_quantum}, and a previous study on the semi-classical quantisation of isotropic case has been performed in Ref.~\cite{Bargheer_2008}. 

In the scope of this article, we discuss the simplest solutions with only one mode number $n_1$ and the semi-classical limit of them is the 1-cut rational solutions, describing the precessional motion of the spin field~\cite{YM_classical_quantum}. We will refer corresponding solutions to asymptotic Bethe equation~\eqref{eq:ABE} as 1-cut solutions.

\section{Exact solution to asymptotic Bethe equation}
\label{sec:solution}

The asymptotic Bethe equation~\eqref{eq:ABE} can be formulated as an equation for zeros of orthogonal polynomials. This insight has been pointed out in~\cite{Shastry_2001} for the 1-cut solution of the isotropic case ($\delta = 0$), in which case the solutions can be expressed as Laguerre polynomials. This approach gives the possibility of obtaining an analytic expression on the distribution of the Bethe roots, e.g. in~\cite{Finkelshtein_2001}. It gives an opportunity to obtain the contour {\it analytically} where the density is supported taking into account the ``reality condition'', i.e. $\rho (\mu) \dd \mu \in \mathbb{R}$ in Eq.~\eqref{eq:contiABE}. More discussions can be found in Ref.~\cite{YM_classical_quantum}.

When $\delta > 0$, the 1-cut solutions to the asymptotic Bethe equation~\eqref{eq:ABE} are the zeros of generalised Jacobi polynomial in the absence of {\it condensate}~\cite{YM_classical_quantum}, which is a manifestation of the divergence of finite-size (quantum) correction~\eqref{eq:finite_size_mu}. We show the derivation as follows. A similar phenomenon also appears when taking the semi-classical limit of attractive Lieb-Liniger model, cf. Ref.~\cite{Flassig_2016, Piroli_2016}.


To begin with, we can rewrite the asymptotic Bethe equation~\eqref{eq:ABE} as
\be 
	L \left( \mu_j - \frac{2 \pi n_1}{\ell} \right) = 2 \sum_{k\neq j}^M \frac{1}{\frac{1}{\mu_j} - \frac{1}{\mu_k} } - 2 \delta \sum_{k \neq j}^M \frac{1}{\mu_j - \mu_k} ,
\ee
with the mode number of the branch cut $n_1$.

We can define the following polynomial $\mathcal{P}(\mu)$ with zeros being the solution to asymptotic Bethe equation~\eqref{eq:ABE}, similar to the Baxter's Q-function~\eqref{eq:Qfunc}, i.e.
\be 
	\mathcal{P} (\mu ) = \prod_{j=1}^M \left( \mu - \mu_j \right) ,
\ee
and accordingly,
\be 
	\tilde{\mathcal{P} } (\mu ) = \prod_{j=1}^M \left( \frac{1}{\mu} - \frac{1}{\mu_j} \right) = \frac{\mathcal{P} (\mu )}{(-1)^M \mu^M \prod_{j=1}^M \mu_j }.
\ee

One can show that the polynomial above satisfies the following identities,
\be 
	2 \sum_{j\neq k} \frac{1}{\mu_j - \mu_k} = \left. \frac{d^2 \mathcal{P}}{d \mu^2} \left/ \frac{d \mathcal{P}}{d \mu} \right. \right|_{\mu = \mu_j} ,
\ee
and
\be 
	2 \sum_{j\neq k} \frac{1}{\frac{1}{\mu_j} - \frac{1}{\mu_k} } = \left. \frac{d^2 \tilde{\mathcal{P} }}{d (1/ \mu)^2} \left/ \frac{d \tilde{\mathcal{P} }}{d (1/ \mu)} \right. \right|_{\mu = \mu_j} .
\ee

Let us denote $\mathcal{P}^{\prime} (\mu) = \frac{d}{d \mu} \mathcal{P} (\mu)$, and we have
\be 
	\frac{d \tilde{P}}{d (1/ \mu)} = \frac{d \mu }{d (1/\mu )} \frac{d}{d \mu} \left( \frac{\mathcal{P} (\mu)}{(-1)^M \mu^M \prod_{j=1}^M \mu_j } \right)= \frac{M \mu \mathcal{P} (\mu) - \mu^2 \mathcal{P}^\prime (\mu )}{(-1)^M \mu^M \prod_{j=1}^M \mu_j } ,
	\label{eq:dev1}
\ee
and
\be 
	\frac{d^2 \tilde{\mathcal{P} }}{d (1/ \mu)^2} = \frac{M(M-1) \mu^2 \mathcal{P} (\mu) - 2 (M-1) \mu^3 \mathcal{P}^\prime (\mu ) + \mu^4 \mathcal{P}^{\prime \prime} (\mu )}{(-1)^M \mu^M \prod_{j=1}^M \mu_j } .
	\label{eq:dev2}
\ee

Using Eqs.~\eqref{eq:dev1} and \eqref{eq:dev2}, we can rewrite the asymptotic Bethe equation~\eqref{eq:ABE} in terms of polynomial $\mathcal{P} (\mu )$, i.e.
\be 
	L \left(\mu_j - \frac{2 \pi n_1 }{\ell} \right) \mathcal{P}^\prime (\mu_j) = 2(M-1) \mu_j \mathcal{P}^\prime (\mu_j) - \mu_j^2 \mathcal{P}^{\prime \prime}(\mu_j) - \delta \mathcal{P}^{\prime \prime} (\mu_j) ,
\ee
or equivalently
\be 
	(\mu_j^2 + \delta ) \mathcal{P}^{\prime \prime} (\mu_j )+ \left[ (L - 2 M +2) \mu_j - \frac{2 \pi n_1 }{\ell} L \right] \mathcal{P}^\prime (\mu_j) = 0 \propto \mathcal{P}(\mu_j) , \quad \forall j = 1 , \cdots M .
\ee

By matching the coefficient of the highest order, we obtain the following ordinary differential equation that the polynomial $\mathcal{P} (\mu )$ satisfies,
\be 
	(\mu^2 + \delta) \mathcal{P}^{\prime \prime} (\mu ) + \left[ (L - 2M +2) \mu - \frac{2 \pi n_1 }{\ell} L \right] \mathcal{P}^\prime (\mu) - M(L -M+1) \mathcal{P}(\mu) = 0 .
\ee

The solutions to this ordinary differential equation can be expressed in terms of the renowned generalised Jacobi polynomial $P_{K}^{(a,b)} (x)$. \footnote{The reason why it is called ``generalised'' is that parameters $a$ and $b$ are complex in our case rather than $a,b>-1$ in the conventional case.}

Generalised Jacobi polynomial satisfies the following ordinary differential equation~\cite{Szego_1959},
\be 
\begin{split}
	& (1 - y^2) P^{\prime \prime} (y) + [b - a - (a+b+2) y] P^\prime (y) + K(K+a+b+1) P (y) = 0 , \\ 
	& P (y) = P^{(a,b)}_K (y) .
\end{split}
\ee

Through a simple change of variable $z = \mu / (i \sqrt{\delta}) $, we have
\be 
	(z^2 - 1) \mathsf{P}^{\prime \prime} (z ) + \left[ (L - 2 M +2) z - \frac{2 \pi n_1}{i \sqrt{\delta}  \ell} L \right] \mathsf{P}^\prime (z) - M(L-M+1) \mathsf{P}(z) = 0 ,
\ee
with $\mathsf{P} (z) = \prod_{j=1}^M (z-z_j)$. The following identification can be found by matching the coefficients,
\be 
	a - b = - \frac{2 \pi n_1}{i \sqrt{\delta} \ell} L , \quad a + b = L - 2 M , \quad K = M .	
\ee

Therefore, the (polynomial) solution to the original ordinary differential equation is
\be 
	\mathcal{P} (\mu) \propto P_M^{a,b} \left( \frac{ \mu}{i \sqrt{\delta} } \right) .
	\label{eq:exactsol}
\ee

Therefore, the 1-cut solution to the asymptotic Bethe equation~\eqref{eq:ABE} coincide with the zeros of the corresponding Jacobi polynomial divided by $i \sqrt{\delta}$. One might notice that it is singular to take the isotropic limit $\delta \to 0$. Indeed, it is not straightforward to take the limit $a-b \to 0$ of the generalised Jacobi polynomial. The correct approach is to start with asymptotic Bethe equation in terms of spectral parameter $\zeta = 1/\mu $ and take the isotropic limit directly so that the solution to it can be expressed as zeros of Laguerre polynomial as shown in~\cite{Shastry_2001}. We demonstrate the procedure in Appendix~\ref{app:isotropic}.

\begin{figure}
\centering
\includegraphics[width=.7\linewidth]{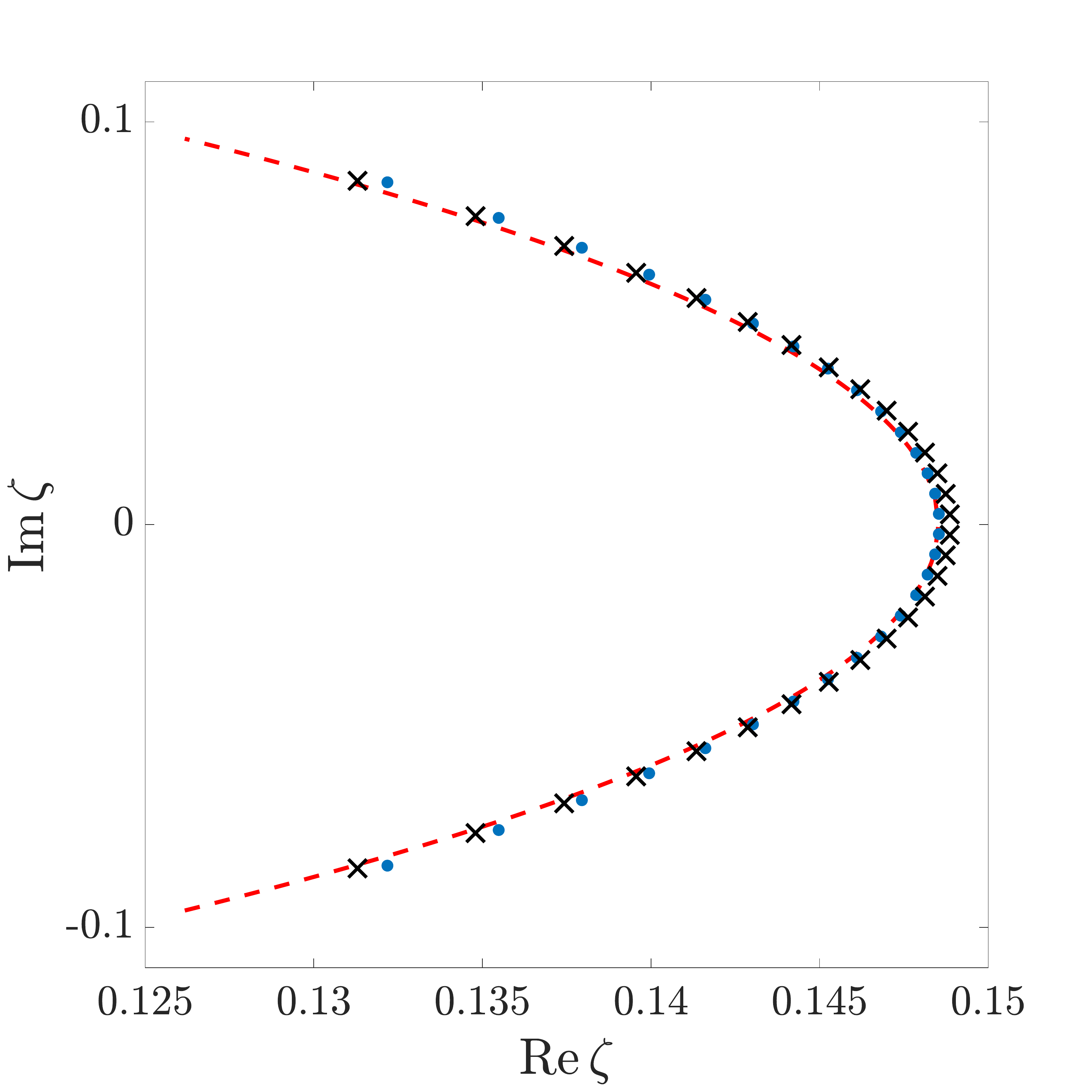}
\caption{The comparison between solution to Eq.~\eqref{eq:exactsol}, the numerical solution to a finite-size Bethe equation and the branch cut for $L=300$, $M=30$, $n_1 = 1$. The figure is plotted in $\zeta = 1 / \mu$ variable. The blue dots denote the numerical solution to Bethe equation~\eqref{eq:betheeq}. The black crosses are the solution to Eq.~\eqref{eq:exactsol} and the red dashed line is the location of the branch cut in the limit $L\to \infty$.}
\label{fig:compare}
\end{figure}

\paragraph*{Remarks.} When the finite-size (quantum) correction~\eqref{eq:finite_size_mu} diverges, we need to take into account the Bethe equations~\eqref{eq:betheeq} non-perturbatively. This gives rise to the phenomenon of {\it condensate}~\cite{YM_classical_quantum, Bargheer_2008}, i.e. Bethe roots forming a segment of logarithmic branch cut with constant density. So long as the filling fraction $\nu = M/L$ is not too high, cf. Ref.~\cite{YM_classical_quantum}, we could safely omit the finite-size correction and our result fits well with the numerical results.

\paragraph*{Continuous limit.} In the continuous limit, the semi-classical eigenstates that we describe using the zeros of generalised Jacobi polynomial are the low-energy excitations of ferromagnetic ground state with fixed magnetisation, as described in Ref.~\cite{YM_classical_quantum} in details. We could determine the location of the branch cut numerically using method in Ref.~\cite{YM_classical_quantum}, and the comparison between Eq.~\eqref{eq:exactsol}, the numerical solution to a finite-size Bethe equation and the branch cut is shown in Fig.~\ref{fig:compare}. As we could observe in Fig.~\ref{fig:compare}, the solution to Eq.~\eqref{eq:exactsol} can be considered to be a good approximation to the solution to Bethe equation~\eqref{eq:betheeq} as system size $L$ increases, and in the limit $L\to \infty$, both of them converge to the contour of the branch cut.

\section{Conclusion}
\label{sec:conclusion}

We present an exact solution to the asymptotic Bethe equation to the weakly anisotropic Heisenberg spin chain, i.e. 1-cut solution. In the continuous limit 1-cut solutions become the quantised non-linear spin waves, as shown in Ref.~\cite{YM_classical_quantum}. The exact solution to asymptotic Bethe equation~\eqref{eq:ABE} are the zeros of generalised Jacobi polynomial.

The result makes it possible to obtain the analytic expression of the contour of spectral curve (i.e. the branch cut where the Bethe roots condense) using the information of the distribution of zeros of generalised Jacobi polynomial when taking the $L \to \infty$ limit, using similar method in~\cite{Finkelshtein_2001}. In comparison, at the moment the contour of the branch cut is determined numerically. In addition, it would be interesting to consider more generic scenarios, such as more than 1 branch cuts are present, and the cases with condensate. However, to obtain exact solutions becomes more difficult due to the interactions between cuts, and we leave it for future investigations.

\section*{Acknowledgements}
The author thanks Oleksandr Gamayun and Enej Ilievski for helpful discussions and collaborations. The author acknowledges the support from the European Research Council under ERC Advanced grant 743032 \textsc{dynamint}.

\begin{appendix}

\section{Isotropic limit}
\label{app:isotropic}

At isotropic limit ($\Delta = 1$), the Bethe equation becomes
\be 
	\left( \frac{\lambda_j  + \ii /2 }{\lambda_j - \ii /2} \right)^L \prod_{k \neq j}^M \frac{\lambda_j - \lambda_k - \ii}{\lambda_j - \lambda_k + \ii} = 1 .
\ee

In order to obtain the asymptotic Bethe equation, we need to consider the following scaling limit $x = \frac{\lambda \ell}{L}$, and after taking the logarithmic form and taking the limit $L\to \infty$, we have
\be 
	\frac{\ell}{x_j} = 2 \pi n_j + \frac{2 \ell}{L} \sum_{k\neq j} \frac{1}{x_j - x_k } ,
	\label{eq:isoasympbetheeq}
\ee
i.e. asymptotic Bethe equation in the isotropic limit.

First of all, we assume that we only cope with 1-cut solution, i.e. there is only 1 mode number $n_1$. We introduce a new variable $X$,
\be 
	X = - \frac{2 \pi n_1}{\ell} L x ,
\ee
and define the following polynomial with zeros being the solution of asymptotic Bethe equation~\ref{eq:isoasympbetheeq},
\be 
	\mathsf{P}_{\rm iso} (X) = \prod_{j=1}^M \left( X - X_j \right) .
\ee
Using the method outlined in Section~\ref{sec:solution}, we obtain a differential equation that $\mathsf{P}_{\rm iso} (X)$ satisfies,
\be 
	X \mathsf{P}_{\rm iso}^{\prime \prime} (X) + \left( - L - X \right)  \mathsf{P}_{\rm iso}^{\prime} (X) + M \mathsf{P}_{\rm iso} (X) = 0 .
\ee
The solution can be expressed in terms of Laguerre polynomial $\mathcal{L}^b_a (u)$~\cite{Szego_1959, Dhar_2000}, i.e.
\be 
	 \mathsf{P}_{\rm iso} (X) \propto \mathcal{L}_M^{-L -1} \left( X \right) .
\ee
In terms of $x$, we have
\be 
	\mathcal{P}_{\mathrm{iso}} (x) \propto \mathcal{L}_M^{-L -1} \left( - \frac{2 \pi n_1 }{\ell}L x  \right) , \quad \mathcal{P}_{\mathrm{iso}} (x)  = \prod_{j=1}^M \left( x - x_j \right) . 
\ee

When the filling fraction $\nu$ becomes larger, there is still the problem with the emergence of condensate, cf. Ref.~\cite{Bargheer_2008}.

Since $ \sin \left[ \eta \left(\frac{\vartheta}{\eta} +\frac{\ii}{2} \right) \right] \simeq \eta \left( \lambda+ \frac{\ii}{2} \right)$ when $\eta$ is small, the spectral parameter $x$ can be considered as the isotropic limit of $\zeta = 1/\mu$, as explained in the Section~\ref{sec:solution}. We can rewrite Eq.~\eqref{eq:ABE} in terms of $\zeta$, i.e.
\be 
	\frac{\ell }{\zeta_j} = 2 \pi n_j + \frac{2 \ell }{L} \sum_{k\neq j}^M \frac{1+ \delta \zeta_j \zeta_k}{\zeta_j - \zeta_k} ,
\ee
and taking the limit $\delta \to 0$, we recover the asymptotic Bethe equation at isotropic limit~\eqref{eq:isoasympbetheeq}.

\end{appendix}

\bibliography{bib_draft}

\begin{thebibliography}{27}%
\makeatletter
\providecommand \@ifxundefined [1]{%
 \@ifx{#1\undefined}
}%
\providecommand \@ifnum [1]{%
 \ifnum #1\expandafter \@firstoftwo
 \else \expandafter \@secondoftwo
 \fi
}%
\providecommand \@ifx [1]{%
 \ifx #1\expandafter \@firstoftwo
 \else \expandafter \@secondoftwo
 \fi
}%
\providecommand \natexlab [1]{#1}%
\providecommand \enquote  [1]{``#1''}%
\providecommand \bibnamefont  [1]{#1}%
\providecommand \bibfnamefont [1]{#1}%
\providecommand \citenamefont [1]{#1}%
\providecommand \href@noop [0]{\@secondoftwo}%
\providecommand \href [0]{\begingroup \@sanitize@url \@href}%
\providecommand \@href[1]{\@@startlink{#1}\@@href}%
\providecommand \@@href[1]{\endgroup#1\@@endlink}%
\providecommand \@sanitize@url [0]{\catcode `\\12\catcode `\$12\catcode
  `\&12\catcode `\#12\catcode `\^12\catcode `\_12\catcode `\%12\relax}%
\providecommand \@@startlink[1]{}%
\providecommand \@@endlink[0]{}%
\providecommand \url  [0]{\begingroup\@sanitize@url \@url }%
\providecommand \@url [1]{\endgroup\@href {#1}{\urlprefix }}%
\providecommand \urlprefix  [0]{URL }%
\providecommand \Eprint [0]{\href }%
\providecommand \doibase [0]{http://dx.doi.org/}%
\providecommand \selectlanguage [0]{\@gobble}%
\providecommand \bibinfo  [0]{\@secondoftwo}%
\providecommand \bibfield  [0]{\@secondoftwo}%
\providecommand \translation [1]{[#1]}%
\providecommand \BibitemOpen [0]{}%
\providecommand \bibitemStop [0]{}%
\providecommand \bibitemNoStop [0]{.\EOS\space}%
\providecommand \EOS [0]{\spacefactor3000\relax}%
\providecommand \BibitemShut  [1]{\csname bibitem#1\endcsname}%
\let\auto@bib@innerbib\@empty
\bibitem [{\citenamefont {Gamayun}\ \emph {et~al.}(2019)\citenamefont
  {Gamayun}, \citenamefont {Miao},\ and\ \citenamefont
  {Ilievski}}]{Gamayun_2019}%
  \BibitemOpen
  \bibfield  {author} {\bibinfo {author} {\bibfnamefont {O.}~\bibnamefont
  {Gamayun}}, \bibinfo {author} {\bibfnamefont {Y.}~\bibnamefont {Miao}}, \
  and\ \bibinfo {author} {\bibfnamefont {E.}~\bibnamefont {Ilievski}},\ }\href
  {\doibase 10.1103/PhysRevB.99.140301} {\bibfield  {journal} {\bibinfo
  {journal} {Phys. Rev. B}\ }\textbf {\bibinfo {volume} {99}},\ \bibinfo
  {pages} {140301} (\bibinfo {year} {2019})}\BibitemShut {NoStop}%
\bibitem [{\citenamefont {Misguich}\ \emph {et~al.}(2019)\citenamefont
  {Misguich}, \citenamefont {Pavloff},\ and\ \citenamefont
  {Pasquier}}]{Misguich_2019}%
  \BibitemOpen
  \bibfield  {author} {\bibinfo {author} {\bibfnamefont {G.}~\bibnamefont
  {Misguich}}, \bibinfo {author} {\bibfnamefont {N.}~\bibnamefont {Pavloff}}, \
  and\ \bibinfo {author} {\bibfnamefont {V.}~\bibnamefont {Pasquier}},\ }\href
  {\doibase 10.21468/SciPostPhys.7.2.025} {\bibfield  {journal} {\bibinfo
  {journal} {SciPost Phys.}\ }\textbf {\bibinfo {volume} {7}},\ \bibinfo
  {pages} {25} (\bibinfo {year} {2019})}\BibitemShut {NoStop}%
\bibitem [{\citenamefont {Miao}\ \emph {et~al.}(2020)\citenamefont {Miao},
  \citenamefont {Ilievski},\ and\ \citenamefont
  {Gamayun}}]{YM_classical_quantum}%
  \BibitemOpen
  \bibfield  {author} {\bibinfo {author} {\bibfnamefont {Y.}~\bibnamefont
  {Miao}}, \bibinfo {author} {\bibfnamefont {E.}~\bibnamefont {Ilievski}}, \
  and\ \bibinfo {author} {\bibfnamefont {O.}~\bibnamefont {Gamayun}},\ }\href
  {https://arxiv.org/abs/2010.07232} {\bibfield  {journal} {\bibinfo  {journal}
  {arXiv preprint arXiv:2010.07232}\ } (\bibinfo {year} {2020})}\BibitemShut
  {NoStop}%
\bibitem [{\citenamefont {\v{Z}nidari\v{c}}(2011)}]{Znidaric_2011}%
  \BibitemOpen
  \bibfield  {author} {\bibinfo {author} {\bibfnamefont {M.}~\bibnamefont
  {\v{Z}nidari\v{c}}},\ }\href {\doibase 10.1103/PhysRevLett.106.220601}
  {\bibfield  {journal} {\bibinfo  {journal} {Phys. Rev. Lett.}\ }\textbf
  {\bibinfo {volume} {106}},\ \bibinfo {pages} {220601} (\bibinfo {year}
  {2011})}\BibitemShut {NoStop}%
\bibitem [{\citenamefont {Ljubotina}\ \emph {et~al.}(2017)\citenamefont
  {Ljubotina}, \citenamefont {{\v{Z}}nidari{\v{c}}},\ and\ \citenamefont
  {Prosen}}]{Ljubotina_2017}%
  \BibitemOpen
  \bibfield  {author} {\bibinfo {author} {\bibfnamefont {M.}~\bibnamefont
  {Ljubotina}}, \bibinfo {author} {\bibfnamefont {M.}~\bibnamefont
  {{\v{Z}}nidari{\v{c}}}}, \ and\ \bibinfo {author} {\bibfnamefont
  {T.}~\bibnamefont {Prosen}},\ }\href {\doibase 10.1038/ncomms16117}
  {\bibfield  {journal} {\bibinfo  {journal} {Nat. Comm.}\ }\textbf {\bibinfo
  {volume} {8}} (\bibinfo {year} {2017}),\ 10.1038/ncomms16117}\BibitemShut
  {NoStop}%
\bibitem [{\citenamefont {Ilievski}\ \emph {et~al.}(2018)\citenamefont
  {Ilievski}, \citenamefont {{De Nardis}}, \citenamefont {Medenjak},\ and\
  \citenamefont {Prosen}}]{Ilievski_2018}%
  \BibitemOpen
  \bibfield  {author} {\bibinfo {author} {\bibfnamefont {E.}~\bibnamefont
  {Ilievski}}, \bibinfo {author} {\bibfnamefont {J.}~\bibnamefont {{De
  Nardis}}}, \bibinfo {author} {\bibfnamefont {M.}~\bibnamefont {Medenjak}}, \
  and\ \bibinfo {author} {\bibfnamefont {T.}~\bibnamefont {Prosen}},\ }\href
  {\doibase 10.1103/PhysRevLett.121.230602} {\bibfield  {journal} {\bibinfo
  {journal} {Phys. Rev. Lett.}\ }\textbf {\bibinfo {volume} {121}},\ \bibinfo
  {pages} {230602} (\bibinfo {year} {2018})}\BibitemShut {NoStop}%
\bibitem [{\citenamefont {{De Nardis}}\ \emph {et~al.}(2019)\citenamefont {{De
  Nardis}}, \citenamefont {Medenjak}, \citenamefont {Karrasch},\ and\
  \citenamefont {Ilievski}}]{De_Nardis_2019}%
  \BibitemOpen
  \bibfield  {author} {\bibinfo {author} {\bibfnamefont {J.}~\bibnamefont {{De
  Nardis}}}, \bibinfo {author} {\bibfnamefont {M.}~\bibnamefont {Medenjak}},
  \bibinfo {author} {\bibfnamefont {C.}~\bibnamefont {Karrasch}}, \ and\
  \bibinfo {author} {\bibfnamefont {E.}~\bibnamefont {Ilievski}},\ }\href
  {\doibase 10.1103/PhysRevLett.123.186601} {\bibfield  {journal} {\bibinfo
  {journal} {Phys. Rev. Lett.}\ }\textbf {\bibinfo {volume} {123}},\ \bibinfo
  {pages} {186601} (\bibinfo {year} {2019})}\BibitemShut {NoStop}%
\bibitem [{\citenamefont {Gopalakrishnan}\ and\ \citenamefont
  {Vasseur}(2019)}]{Vasseur_2019}%
  \BibitemOpen
  \bibfield  {author} {\bibinfo {author} {\bibfnamefont {S.}~\bibnamefont
  {Gopalakrishnan}}\ and\ \bibinfo {author} {\bibfnamefont {R.}~\bibnamefont
  {Vasseur}},\ }\href {\doibase 10.1103/PhysRevLett.122.127202} {\bibfield
  {journal} {\bibinfo  {journal} {Phys. Rev. Lett.}\ }\textbf {\bibinfo
  {volume} {122}},\ \bibinfo {pages} {127202} (\bibinfo {year}
  {2019})}\BibitemShut {NoStop}%
\bibitem [{\citenamefont {{De Nardis}}\ \emph
  {et~al.}(2020{\natexlab{a}})\citenamefont {{De Nardis}}, \citenamefont
  {Medenjak}, \citenamefont {Karrasch},\ and\ \citenamefont
  {Ilievski}}]{De_Nardis_2020}%
  \BibitemOpen
  \bibfield  {author} {\bibinfo {author} {\bibfnamefont {J.}~\bibnamefont {{De
  Nardis}}}, \bibinfo {author} {\bibfnamefont {M.}~\bibnamefont {Medenjak}},
  \bibinfo {author} {\bibfnamefont {C.}~\bibnamefont {Karrasch}}, \ and\
  \bibinfo {author} {\bibfnamefont {E.}~\bibnamefont {Ilievski}},\ }\href
  {\doibase 10.1103/PhysRevLett.124.210605} {\bibfield  {journal} {\bibinfo
  {journal} {Phys. Rev. Lett.}\ }\textbf {\bibinfo {volume} {124}},\ \bibinfo
  {pages} {210605} (\bibinfo {year} {2020}{\natexlab{a}})}\BibitemShut
  {NoStop}%
\bibitem [{\citenamefont {{De Nardis}}\ \emph
  {et~al.}(2020{\natexlab{b}})\citenamefont {{De Nardis}}, \citenamefont
  {Gopalakrishnan}, \citenamefont {Ilievski},\ and\ \citenamefont
  {Vasseur}}]{De_Nardis_2020_2}%
  \BibitemOpen
  \bibfield  {author} {\bibinfo {author} {\bibfnamefont {J.}~\bibnamefont {{De
  Nardis}}}, \bibinfo {author} {\bibfnamefont {S.}~\bibnamefont
  {Gopalakrishnan}}, \bibinfo {author} {\bibfnamefont {E.}~\bibnamefont
  {Ilievski}}, \ and\ \bibinfo {author} {\bibfnamefont {R.}~\bibnamefont
  {Vasseur}},\ }\href {\doibase 10.1103/PhysRevLett.125.070601} {\bibfield
  {journal} {\bibinfo  {journal} {Phys. Rev. Lett.}\ }\textbf {\bibinfo
  {volume} {125}},\ \bibinfo {pages} {070601} (\bibinfo {year}
  {2020}{\natexlab{b}})}\BibitemShut {NoStop}%
\bibitem [{\citenamefont {Sutherland}(1978)}]{Sutherland_1978}%
  \BibitemOpen
  \bibfield  {author} {\bibinfo {author} {\bibfnamefont {B.}~\bibnamefont
  {Sutherland}},\ }\href {http://www.jstor.org/stable/44239277} {\bibfield
  {journal} {\bibinfo  {journal} {Rocky Mt. J. Math.}\ }\textbf {\bibinfo
  {volume} {8}},\ \bibinfo {pages} {413} (\bibinfo {year} {1978})}\BibitemShut
  {NoStop}%
\bibitem [{\citenamefont {Sutherland}(1995)}]{Sutherland_1995}%
  \BibitemOpen
  \bibfield  {author} {\bibinfo {author} {\bibfnamefont {B.}~\bibnamefont
  {Sutherland}},\ }\href {\doibase 10.1103/PhysRevLett.74.816} {\bibfield
  {journal} {\bibinfo  {journal} {Phys. Rev. Lett.}\ }\textbf {\bibinfo
  {volume} {74}},\ \bibinfo {pages} {816} (\bibinfo {year} {1995})}\BibitemShut
  {NoStop}%
\bibitem [{\citenamefont {Dhar}\ and\ \citenamefont
  {Sriram~Shastry}(2000)}]{Dhar_2000}%
  \BibitemOpen
  \bibfield  {author} {\bibinfo {author} {\bibfnamefont {A.}~\bibnamefont
  {Dhar}}\ and\ \bibinfo {author} {\bibfnamefont {B.}~\bibnamefont
  {Sriram~Shastry}},\ }\href {\doibase 10.1103/PhysRevLett.85.2813} {\bibfield
  {journal} {\bibinfo  {journal} {Phys. Rev. Lett.}\ }\textbf {\bibinfo
  {volume} {85}},\ \bibinfo {pages} {2813} (\bibinfo {year}
  {2000})}\BibitemShut {NoStop}%
\bibitem [{\citenamefont {Kazakov}\ \emph {et~al.}(2004)\citenamefont
  {Kazakov}, \citenamefont {Marshakov}, \citenamefont {Minahan},\ and\
  \citenamefont {Zarembo}}]{Kazakov_2004}%
  \BibitemOpen
  \bibfield  {author} {\bibinfo {author} {\bibfnamefont {V.~A.}\ \bibnamefont
  {Kazakov}}, \bibinfo {author} {\bibfnamefont {A.}~\bibnamefont {Marshakov}},
  \bibinfo {author} {\bibfnamefont {J.~A.}\ \bibnamefont {Minahan}}, \ and\
  \bibinfo {author} {\bibfnamefont {K.}~\bibnamefont {Zarembo}},\ }\href
  {\doibase 10.1088/1126-6708/2004/05/024} {\bibfield  {journal} {\bibinfo
  {journal} {J. High Energy Phys.}\ }\textbf {\bibinfo {volume} {2004}},\
  \bibinfo {pages} {024} (\bibinfo {year} {2004})}\BibitemShut {NoStop}%
\bibitem [{\citenamefont {Bargheer}\ \emph {et~al.}(2008)\citenamefont
  {Bargheer}, \citenamefont {Beisert},\ and\ \citenamefont
  {Gromov}}]{Bargheer_2008}%
  \BibitemOpen
  \bibfield  {author} {\bibinfo {author} {\bibfnamefont {T.}~\bibnamefont
  {Bargheer}}, \bibinfo {author} {\bibfnamefont {N.}~\bibnamefont {Beisert}}, \
  and\ \bibinfo {author} {\bibfnamefont {N.}~\bibnamefont {Gromov}},\ }\href
  {\doibase 10.1088/1367-2630/10/10/103023} {\bibfield  {journal} {\bibinfo
  {journal} {New J. Phys.}\ }\textbf {\bibinfo {volume} {10}},\ \bibinfo
  {pages} {103023} (\bibinfo {year} {2008})}\BibitemShut {NoStop}%
\bibitem [{\citenamefont {Minahan}\ and\ \citenamefont
  {Zarembo}(2003)}]{Minahan_2003}%
  \BibitemOpen
  \bibfield  {author} {\bibinfo {author} {\bibfnamefont {J.~A.}\ \bibnamefont
  {Minahan}}\ and\ \bibinfo {author} {\bibfnamefont {K.}~\bibnamefont
  {Zarembo}},\ }\href {\doibase 10.1088/1126-6708/2003/03/013} {\bibfield
  {journal} {\bibinfo  {journal} {J. High Energy Phys.}\ }\textbf {\bibinfo
  {volume} {2003}},\ \bibinfo {pages} {013} (\bibinfo {year}
  {2003})}\BibitemShut {NoStop}%
\bibitem [{\citenamefont {Kruczenski}(2004)}]{Kruczenski_2004}%
  \BibitemOpen
  \bibfield  {author} {\bibinfo {author} {\bibfnamefont {M.}~\bibnamefont
  {Kruczenski}},\ }\href {\doibase 10.1103/PhysRevLett.93.161602} {\bibfield
  {journal} {\bibinfo  {journal} {Phys. Rev. Lett.}\ }\textbf {\bibinfo
  {volume} {93}},\ \bibinfo {pages} {161602} (\bibinfo {year}
  {2004})}\BibitemShut {NoStop}%
\bibitem [{\citenamefont {Escobedo}\ \emph {et~al.}(2011)\citenamefont
  {Escobedo}, \citenamefont {Gromov}, \citenamefont {Sever},\ and\
  \citenamefont {Vieira}}]{Escobedo_2011}%
  \BibitemOpen
  \bibfield  {author} {\bibinfo {author} {\bibfnamefont {J.}~\bibnamefont
  {Escobedo}}, \bibinfo {author} {\bibfnamefont {N.}~\bibnamefont {Gromov}},
  \bibinfo {author} {\bibfnamefont {A.}~\bibnamefont {Sever}}, \ and\ \bibinfo
  {author} {\bibfnamefont {P.}~\bibnamefont {Vieira}},\ }\href {\doibase
  10.1007/jhep09(2011)028} {\bibfield  {journal} {\bibinfo  {journal} {J. High
  Energy Phys.}\ }\textbf {\bibinfo {volume} {2011}} (\bibinfo {year} {2011}),\
  10.1007/jhep09(2011)028}\BibitemShut {NoStop}%
\bibitem [{\citenamefont {Gromov}\ \emph {et~al.}(2012)\citenamefont {Gromov},
  \citenamefont {Sever},\ and\ \citenamefont {Vieira}}]{Gromov_2012}%
  \BibitemOpen
  \bibfield  {author} {\bibinfo {author} {\bibfnamefont {N.}~\bibnamefont
  {Gromov}}, \bibinfo {author} {\bibfnamefont {A.}~\bibnamefont {Sever}}, \
  and\ \bibinfo {author} {\bibfnamefont {P.}~\bibnamefont {Vieira}},\ }\href
  {\doibase 10.1007/jhep07(2012)044} {\bibfield  {journal} {\bibinfo  {journal}
  {J. High Energy Phys.}\ }\textbf {\bibinfo {volume} {2012}} (\bibinfo {year}
  {2012}),\ 10.1007/jhep07(2012)044}\BibitemShut {NoStop}%
\bibitem [{\citenamefont {Kostov}(2012{\natexlab{a}})}]{Kostov_2012_PRL}%
  \BibitemOpen
  \bibfield  {author} {\bibinfo {author} {\bibfnamefont {I.}~\bibnamefont
  {Kostov}},\ }\href {\doibase 10.1103/PhysRevLett.108.261604} {\bibfield
  {journal} {\bibinfo  {journal} {Phys. Rev. Lett.}\ }\textbf {\bibinfo
  {volume} {108}},\ \bibinfo {pages} {261604} (\bibinfo {year}
  {2012}{\natexlab{a}})}\BibitemShut {NoStop}%
\bibitem [{\citenamefont {Kostov}(2012{\natexlab{b}})}]{Kostov_2012}%
  \BibitemOpen
  \bibfield  {author} {\bibinfo {author} {\bibfnamefont {I.}~\bibnamefont
  {Kostov}},\ }\href {\doibase 10.1088/1751-8113/45/49/494018} {\bibfield
  {journal} {\bibinfo  {journal} {J. Phys. A}\ }\textbf {\bibinfo {volume}
  {45}},\ \bibinfo {pages} {494018} (\bibinfo {year}
  {2012}{\natexlab{b}})}\BibitemShut {NoStop}%
\bibitem [{\citenamefont {Shastry}\ and\ \citenamefont
  {Dhar}(2001)}]{Shastry_2001}%
  \BibitemOpen
  \bibfield  {author} {\bibinfo {author} {\bibfnamefont {B.~S.}\ \bibnamefont
  {Shastry}}\ and\ \bibinfo {author} {\bibfnamefont {A.}~\bibnamefont {Dhar}},\
  }\href {\doibase 10.1088/0305-4470/34/31/313} {\bibfield  {journal} {\bibinfo
   {journal} {J. Phys. A}\ }\textbf {\bibinfo {volume} {34}},\ \bibinfo {pages}
  {6197} (\bibinfo {year} {2001})}\BibitemShut {NoStop}%
\bibitem [{\citenamefont {Beisert}\ \emph {et~al.}(2005)\citenamefont
  {Beisert}, \citenamefont {Tseytlin},\ and\ \citenamefont
  {Zarembo}}]{Beisert_2005}%
  \BibitemOpen
  \bibfield  {author} {\bibinfo {author} {\bibfnamefont {N.}~\bibnamefont
  {Beisert}}, \bibinfo {author} {\bibfnamefont {A.~A.}\ \bibnamefont
  {Tseytlin}}, \ and\ \bibinfo {author} {\bibfnamefont {K.}~\bibnamefont
  {Zarembo}},\ }\href {\doibase 10.1016/j.nuclphysb.2005.03.030} {\bibfield
  {journal} {\bibinfo  {journal} {Nucl. Phys. B.}\ }\textbf {\bibinfo {volume}
  {715}},\ \bibinfo {pages} {190} (\bibinfo {year} {2005})}\BibitemShut
  {NoStop}%
\bibitem [{\citenamefont {Mart{\'i}nez-Finkelshtein}\ \emph
  {et~al.}(2001)\citenamefont {Mart{\'i}nez-Finkelshtein}, \citenamefont
  {Mart{\'i}nez-Gonz{\'{a}}lez},\ and\ \citenamefont
  {Orive}}]{Finkelshtein_2001}%
  \BibitemOpen
  \bibfield  {author} {\bibinfo {author} {\bibfnamefont {A.}~\bibnamefont
  {Mart{\'i}nez-Finkelshtein}}, \bibinfo {author} {\bibfnamefont
  {P.}~\bibnamefont {Mart{\'i}nez-Gonz{\'{a}}lez}}, \ and\ \bibinfo {author}
  {\bibfnamefont {R.}~\bibnamefont {Orive}},\ }\href {\doibase
  10.1016/s0377-0427(00)00654-3} {\bibfield  {journal} {\bibinfo  {journal} {J.
  Comput. Appl. Math.}\ }\textbf {\bibinfo {volume} {133}},\ \bibinfo {pages}
  {477} (\bibinfo {year} {2001})}\BibitemShut {NoStop}%
\bibitem [{\citenamefont {Flassig}\ \emph {et~al.}(2016)\citenamefont
  {Flassig}, \citenamefont {Franca},\ and\ \citenamefont
  {Pritzel}}]{Flassig_2016}%
  \BibitemOpen
  \bibfield  {author} {\bibinfo {author} {\bibfnamefont {D.}~\bibnamefont
  {Flassig}}, \bibinfo {author} {\bibfnamefont {A.}~\bibnamefont {Franca}}, \
  and\ \bibinfo {author} {\bibfnamefont {A.}~\bibnamefont {Pritzel}},\ }\href
  {\doibase 10.1103/PhysRevA.93.013627} {\bibfield  {journal} {\bibinfo
  {journal} {Phys. Rev. A}\ }\textbf {\bibinfo {volume} {93}},\ \bibinfo
  {pages} {013627} (\bibinfo {year} {2016})}\BibitemShut {NoStop}%
\bibitem [{\citenamefont {Piroli}\ and\ \citenamefont
  {Calabrese}(2016)}]{Piroli_2016}%
  \BibitemOpen
  \bibfield  {author} {\bibinfo {author} {\bibfnamefont {L.}~\bibnamefont
  {Piroli}}\ and\ \bibinfo {author} {\bibfnamefont {P.}~\bibnamefont
  {Calabrese}},\ }\href {\doibase 10.1103/PhysRevA.94.053620} {\bibfield
  {journal} {\bibinfo  {journal} {Phys. Rev. A}\ }\textbf {\bibinfo {volume}
  {94}},\ \bibinfo {pages} {053620} (\bibinfo {year} {2016})}\BibitemShut
  {NoStop}%
\bibitem [{\citenamefont {Szeg{\H{o}}}(1959)}]{Szego_1959}%
  \BibitemOpen
  \bibfield  {author} {\bibinfo {author} {\bibfnamefont {G.}~\bibnamefont
  {Szeg{\H{o}}}},\ }\href@noop {} {\emph {\bibinfo {title} {{Orthogonal
  Polynomials}}}}\ (\bibinfo  {publisher} {American Mathematical Society},\
  \bibinfo {year} {1959})\BibitemShut {NoStop}%
\end{thebibliography}%
\bibliographystyle{apsrev4-1}

\end{document}